\documentstyle[epsfig,macros]{prhep97}
\makeatletter
\let\chapter\hid@chapter
\makeatother
\begin{document}


\authorrunning{F. Csikor\,Z. Fodor}
\titlerunning{{\talknumber}: Excluding light gluinos }
 


\def\talknumber{1104} 
\title{
Closing the light gluino window$^1$ 
}
\author{%
{F. Csikor $ ^\dagger$
(csikor@hal9000.elte.hu) \\
\institute{ Inst. Theor. Phys., E\"otv\"os University\\ }}
{Z. Fodor $ ^{\dagger \dagger}$
\thanks{On leave from Inst. Theor. Phys., E\"otv\"os University}
(fodor@theory4.kek.jp)\\
\institute{ $ ^\dagger$ Inst. Theor. Phys., E\"otv\"os University, Budapest, 
Hungary\\
$ ^{\dagger \dagger}$ KEK, Theory Group, 1--1 Oho, Tsukuba 305, Japan}}}
\maketitle
\vspace{-5.5cm}

{ \normalsize
\hfill \parbox[t]{4cm}{ITP-Budapest 538 \\KEK-TH-551\\
                  } \\[7em]

\addtocounter{footnote}{1}
\footnotetext{
Presented by F. Csikor at the {\it International Europhysics Conference
on High Energy Physics}, Jerusalem, August 19--26 1997.
}

\vspace{2cm}

\begin{abstract}
The running of the strong coupling constant,
$R_{e^+e^-},R_Z$ and $R_\tau$ is studied on the three-loop level. 
Based on experimental data of $R_{e^+e^-},R_Z$
and $R_\tau$ and the LEP multijet analysis, the light gluino scenario is
excluded to 99.97\% CL (window I) and 99.89\% CL (window III).
\end{abstract}
\section{Introduction and motivation}

Asymptotic freedom is one of the most interesting predictions of QCD. 
One can study this by measuring the running coupling constant at different 
energies and comparing the results. Fig. 1 shows $\alpha_s$ running as 
obtained from different experiments together with QCD predictions. 
Of course QCD is in very good shape. Nevertheless one may quantify this 
statement.  This is not the subject of the present talk, for details 
see \cite{r1}.  One may also 
speculate on slower running than required by QCD. A popular possibility is 
the light gluino extension of QCD. In this paper we discuss that this 
scenario can be excluded using three-loop perturbative results and the 
existing experimental data \cite{r1}. 

\begin{figure}
\begin{center}
\epsfig{file=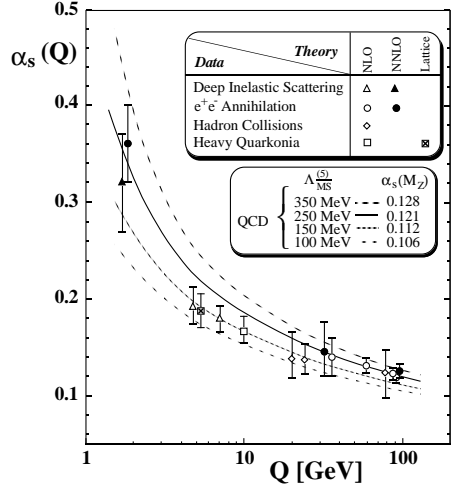,width=6cm}
\caption{$\alpha_s$ running and QCD predictions}
\end{center}
\end{figure}

The experimentally excluded regions of light gluinos (1996 status) 
-- mass v. lifetime -- are shown in Fig. 2.  The moral is that window I., 
i.e. lighter than 1.5 GeV and window III. i.e. masses between 3 GeV and 5 
GeV are allowed by these results.

Last year has brought important technical development of the subject, 
namely 3-loop results have been calculated \cite{r5}--\cite{r7}. It has been 
emphasized that light gluinos are obtained naturally from gauge mediated 
SUSY breaking \cite{r3,r4}. Also some new papers 
both experimental \cite{r8}--\cite{r10} and theoretical  
\cite{r11,r12} presenting 
results on light gluino exclusion have appeared. \cite{r13}--\cite{r15} 
critisizes the exclusion results.

\section{Theoretical and experimental inputs}

Describing $R_{e^+e^-},R_Z$ and $R_\tau$  in the light gluino scenario 
qualitative differences compared to ordinary QCD appear. Since the 
effective number of fermions is larger, running of $\alpha_s(Q)$ becomes 
 slower. 
Including the new degrees of freedom in cross section calculations and 
taking into account the larger phase-space one finds that a smaller $\alpha_s$ 
describes the data at any given energy. 

\begin{figure}
\begin{center}
\epsfig{file=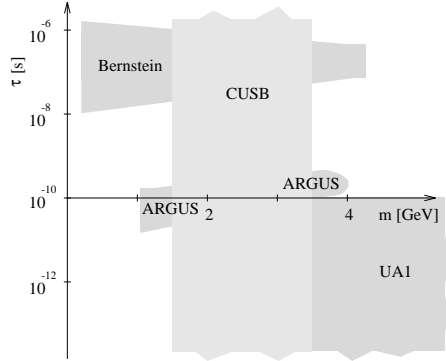,width=6cm}
\caption{Light gluino exclusion limits}
\end{center}
\end{figure}

The basic idea of the present paper is the following.
We suppose that strong interaction is described by a gauge theory based 
on a simple Lie-group, thus hadronic cross-sections, widths, $\beta$-function 
are given by $C_F,C_A,T_F$ and the active number of fermions. Then by looking 
for $C_F,C_A,T_F,n_{\tilde g}$ to describe data accurately, we can determine 
the CL a given gauge theory is excluded or supported by data. 
Since the energy range from $M_\tau$ to $M_Z$ is rather large, it is 
nontrivial for a hypothetical theory to reproduce both the cross section 
values and the running of $\alpha_s(Q)$.

The theoretical inputs of our analyses are ${\cal O}(\alpha_s^3)$ 
calculations of $R_{e^+e^-},R_Z$ and $R_\tau$ for arbitrary group theoretical 
coefficients. The experimental inputs of the analysis are 
$R_\tau$ given by ALEPH and CLEO $R_\tau=3.616(20)$),  
$R_Z$ given by the LEP groups ($R_Z=20.778(29)$) and 
hadronic cross sections at energies 5 GeV--$M_Z$ (all existing published 
data (and some unpublished)). The 
total number of data points included is 182.

\section{Treatment of different sources of errors}

Besides the data the errors are also very important for a fit. Being 
too optomistic destroys reliability of the results. When realized 
that data imply gluino exclusion, we have choosen to use very 
conservative error estimates. 

We treat uncertainties in a unified manner, add the systematic errors linearly 
and total systematic and statistical errors quadratically. The accuracy of 
$R_{e^+e^-}$ and $R_Z$ is  limited by experiments, while for 
$R_\tau$  theoretical uncertainty dominates. For higher order perturbative 
corrections we suppose that the error is the last computed term (asymptotic 
series). For $R_\tau$ this gives significantly larger error than usually 
assumed. For $R_\tau$ mass and nonperturbative corrections are taken into 
account following \cite{r16}. 

Since experimental errors are correlated, we minimize 
\begin{equation}
\chi^2=\Delta^TV^{-1}\Delta \enspace .\label{k1}
\end{equation}
$\Delta$ is an $n$-vector of the residuals of $R_i-R_{fit}$,
$n$ is the number of individual results,
V is $n\times n$ error matrix.

\section{Light gluino results from running of $\alpha_s (Q)$}

First we present results from fiting $R_{e^+e^-}$, $R_Z$, $R_\tau$, i.e. 
$\alpha_s$ running. An example of a two parameter ($x=C_A/C_F , y=T_f /C_F $) 
fit is shown in Fig. 3 for the case of window III. More quantitatively: 
we exclude window III light gluinos with a small mass dependence to 93(91)\% CL 
for $m_{\tilde g}$=3(5) GeV. For window I we do not use  $R_\tau$ in the fit 
and get only 71 \% exclusion CL.

\begin{figure}
\begin{center}
\epsfig{file=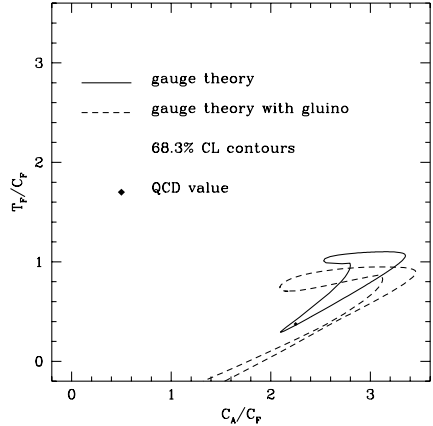,width=6cm}
\caption{Exclusion from $\alpha_s $ running for $m_{\tilde g}$=3 GeV (window 
III)}
\end{center}
\end{figure}

Following \cite{r11} we fixed the underlying group to SU(3) 
and determined the number of gluinos. This corresponds to a one parameter fit
for $n_{\tilde g}$.
For window III we get $n_{\tilde g}$=0.0078$\pm $0.52, 88\% CL
exclusion and for window I $n_{\tilde g}$=-0.070$\pm$0.7,  80\% CL 
exclusion (without $R_\tau$).

Farrar \cite{r13} claims  that increasing our $R_\tau$ error estimate   
 by a factor 2 leads 
to 68\% exlusion for window III case. Performing the calculation 
we get 87(85)\% and 81\% for window III 2 and 1 parameter fits, respectively.

\section{Combination with jet analysis at LEP} 

\begin{figure}
\begin{center}
\epsfig{file=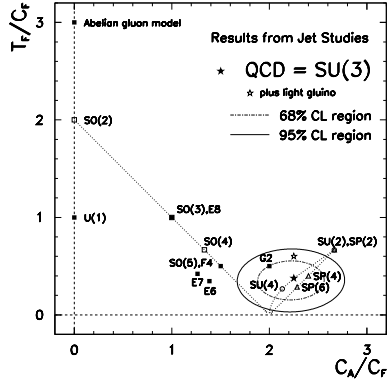,width=6cm}
\caption{LEP 4 jet analyses exclusion results}
\end{center}
\end{figure}

LEP groups have determined the group theoretical factors x,y 
comparing experimental results with leading order 4 jet and higher order
2,3 jet predictions. A summary of the earlier results was given in \cite{r17} 
and shown in Fig. 4. 
To illustrate what happens when our analysis based on $\alpha_s$ running is 
combined with a LEP jet analysis, Fig. 5 shows the 1$\sigma$ 
excluded regions for the OPAL jet analysis alone \cite{r18} (which actually 
does not exclude light gluinos) and ours (copied from Fig. 3). Note that 
the overlap of the two regions is rather small, {\em resulting in stronger 
exclusion than any of the uncombined analyses}. It is also clear that 
increasing the jet analysis ellipsis or shifting it does not change the 
overlap region too much, i.e. the exclusion changes only very little. 

\begin{figure}
\begin{center}
\epsfig{file=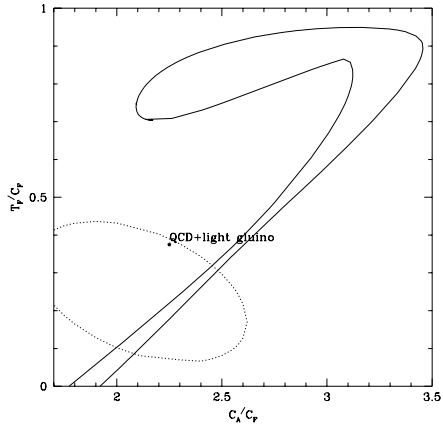,width=6cm}
\caption{Comparison of excluded regions from $\alpha_s$ running and OPAL 
4 jet analysis}
\end{center}
\end{figure}

 In the actual calculation we have included the older LEP jet  analyses 
and the new ALEPH result \cite{r9}.  
 To take into account higher order corrections of 4 jet QCD 
predictions we have increased 
the axes of the error ellipse by 12\% of the theoretical x and y values 
(relative correction of ${\cal O} (\alpha_s)$). 
This change would destroy the predictive 
power of the (uncombined) multi-jet analysis for old data.

\begin{figure}
\begin{center}
\epsfig{file=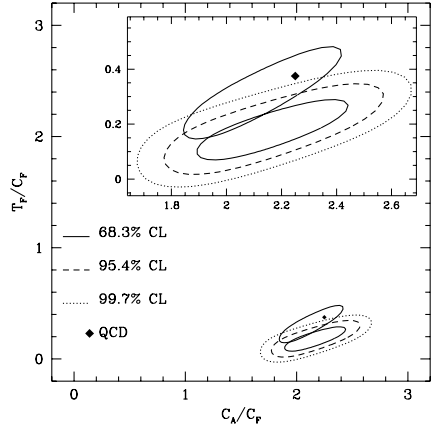,width=6cm}
\caption{Exclusion from combined method, 2 parameter fit}
\end{center}
\end{figure}

An example of the 2  parameter fits is shown in Fig. 6. More quantitatively, 
for window III light gluino ($m_{\tilde g}=$3(5) GeV) the exclusion is 
99.99(99.89)\% CL. For window I  
light gluino  it is: 99.97\% CL.

We also performed 1 parameter fits. 
For window III:
$n_{\tilde g}$=-0.156$\pm$0.27 (-0.197$\pm$0.32)  with 
$m_{\tilde g}=$3(5) GeV i.e. 99.96(99.76)\% CL  exclusion. 
For  window I:
 $n_{\tilde g}$=-0.35$\pm$0.33 i.e. we have 99.96\% CL exclusion.

Farrar \cite{r13} claims that increasing the ALEPH errors by a factor 3 
would lead to 68\% gluino exclusion. Performing the actual calculation
we get at least  95\% exclusion for all cases of the combined analysis.

\section{Summary, conclusion}

We have obtained light gluino exclusion from fits of $R_{e^+e^-},R_Z$ 
and $R_\tau$, i.e. essentally from $\alpha_s$ running. We have combined 
this analysis  with the LEP jet analysis. Our combined analysis  results 
in much more stringent 
light gluino exclusion than LEP jet analysis alone or $\alpha_s$ running alone. The best CL's are  
99.97\% for window I and 99.89\% for window III light gluino exclusion. 
Our results are independent of light gluino lifetime.  \\
This work was partially supported by Hung. Sci.
grants OTKA-T16248,T22929.
 
%

\end{document}